\documentclass[12pt,preprint]{aastex}

\begin{document}

\title{Testing the Disk-Locking Paradigm:\\ An Association Between U-V Excess and Rotation in NGC 2264}

\author{Cassandra Fallscheer and William Herbst}
\affil{Astronomy Department, Wesleyan University, Middletown, CT
06459}
\email{cfallscheer@wesleyan.edu, bill@astro.wesleyan.edu}

\begin{abstract}

We present some results from a UVI photometric study of a field in the young open cluster NGC 2264 aimed, in part, at testing whether accretion in pre-main sequence stars is linked to rotation. We confirm that U-V excess is well correlated with H$\alpha$ equivalent width for the stars in our sample. We show that for the more massive stars in the cluster sample (roughly 0.4-1.2 M$_\sun$) there is also a significant association between U-V excess and rotation, in the sense that slow rotators are more likely to show excess U-band emission and variability. This constitutes significant new evidence in support of the disk-locking paradigm. 

\end{abstract}
 
 
\section{Introduction}

Pre-main sequence (PMS) and zero-age main stars in a solar-like mass range (roughly 0.4-1.2 M$_\sun$) have a bimodal distribution of rotation rates with a very broad range. This fact was discovered by \citet{ah92} in their study of the Orion Nebula Cluster and has been tested, challenged and, ultimately, confirmed by numerous studies in a variety of clusters and associations during the last fifteen years. A recent review is given by \citet{hems06}.

An explanation for slower-than-critical rotation of accreting PMS star was originally proposed by \citet{c90}, \citet{k91} and \citet{s94} and is commonly referred to as the disk-locking mechanism. While various theories differ in detail on exactly how angular momentum is extracted from the system, they agree that a magnetic interaction between the star and an accretion disk is involved. This has led observers to ask whether correlations between accretion disk indicators and stellar rotation exist. The first study to report such evidence was by \citet{e93}. This observational link has also been tested and challenged by many investigators and many studies over the years using different accretion disk proxies such as near-infrared excess emission and H$\alpha$ equivalent width. Among these we cite \citet{h02}, \citet{l04}, \citet{ds05}, \citet{r05}, \citet{n06} and \citet{cb06}. The last two papers present conflicting results on the same cluster, IC 348, and indicate that the controversy is still alive.

It is not our purpose here to critically evaluate any or all of the previous studies but merely to report the results of a new one, designed specifically to test whether stellar rotation and one particular commonly accepted indicator of accretion activity, U-V excess, are associated. Our sample is a set of solar-like PMS stars in the extremely young cluster NGC 2264 with known rotation periods from the studies of \citet{m04} and \citet{l04}. We undertook a UVI photometric study of a field in that cluster from which U-V excesses and U-band variability could be determined. We report a principal result of that study here; a more extensive exposition of results will be published elsewhere.

Solar-like PMS stars may be divided by spectroscopic criteria into so-called classical and weak-line T Tauri stars -- CTTS and WTTS, respectively \citep{s89}. Observational characteristics of accretion activity and accretion disks are stronger in the CTTS, including infrared and ultraviolet excess emission, H$\alpha$ emission and large amplitude, irregular photometric variability. Spectroscopically, the CTTS show strong veiling emission, especially in the blue, and strong H$\alpha$ emission. A common practice is to divide the CTTS and WTTS at an equivalent width of H$\alpha$ equal to 10 \AA. CTTS typically have extremely blue U-V colors because of their veiling and Balmer continuum emission, features which are well modeled as arising from accretion shocks \citep{cg98, g00}. In fact, \citet{g98} have found that the correlation between U-band excess and accretion rates derived from detailed spectral analysis is exceedingly good. In this Letter, we will investigate whether a connection exists between U-V excess, an active accretion indicator, and rotation for a sample of solar-like PMS stars in the extremely young cluster NGC 2264.  

\section {Observations and Sample Selection}

Observations were carried out on ten nights at two separate
 epochs with the S2KB CCD camera on the WIYN 0.9 meter 
telescope at Kitt Peak National Observatory.  Between 8 and 16 January 
2005 a total of six nights of data were obtained. Nearly a year later, four
nights, or partial nights, of data were obtained between 1 and 4 December 2005.  Since we were interested particularly in U-band excesses and variability, most of the data were obtained in U, typically 8 images per night, spread over 6 hours. Some V and I data were obtained each night as well.

The target field is centered at $\alpha$=6h 41m 5.9s  and $\delta$=+9$^{\circ}$ 43$\arcmin$ 6$\arcsec$ (J2000) and includes portions of both of the two principle sub-clusters of TTS in NGC 2264. 
We deliberately avoided the bright star S Mon to the north. Our observing plan was to group observations into 120s and 600s exposures in {\it V},
 a 180s exposure in {\it I}, followed by three 1800s exposures in {\it U}. 
Over the combined observing runs 64, 45, and 28 useful images of NGC 2264 were taken 
through the {\it U}, {\it V}, and {\it I} filters respectively.  We also obtained images of Landolt standard fields \citep{l92} on three photometric nights during the January 2005 observing run for calibration purposes.

All images were initially processed using standard IRAF tasks.  Flat field division was accomplished by
 combining five dome flats taken nightly with each filter.  Using the IRAF routine DAOFIND, a source list of stars in our field was created.  The IRAF task PHOT was used for the photometry with aperture set to 4 pixels.  We chose a sky background annulus inner radius of 15 pixels, and an outer radius of 21 pixels. We obtained data on 960 stars with U magnitudes between 12.5 and 21. Random errors are less than $\sim$0.02 mag for stars brighter than U = 17.5 and increase roughly linearly with magnitude to about 0.25 mag at U = 21.
 
The periodic sample studied here was restricted to the set of stars with R-I $<$ 1.3 (corresponding to V-I $<$ 2.4) that are known periodic variables from the studies of  \citet{m04} and \citet{l04, l05}. All of these stars may confidently be considered members of the NGC 2264 cluster based on those studies. The color restriction assures that we are eliminating lower mass PMS stars, whose rotation properties are known to differ significantly from their higher mass counterparts. The chosen color corresponds to a spectral class of about M2 and an effective temperature of about 3500 K. Models of pre-main sequence evolution assign masses in the range of 0.25-0.4 M$_\odot$ to such stars with more recent models favoring higher values. Roughly speaking the mass range for this sample is, therefore, 0.4 $<$ M/M$_\odot$ $<$ 1.2. The upper limit is set by the rapidly dwindling number of stars with spectral class earlier than K0 for which rotation periods can be determined and could be as high as 1.5 M$_\odot$. 
 
\section{Results}

As a first step in the analysis we identified a set of variable stars by their location in a plot of standard deviation versus average magnitude. A total of 85 stars lying clearly above the non-variable sequence in the U-band was selected from our full set of stars.  In what follows, these variable stars are specially marked. Examination of their light curves and comparison with behavior in V and I shows that there are at least three types of U-band variability. Some stars show rather small amplitudes (0.7 mag or less) and consistent behavior in all colors on time scales of days, indicative of a rotating photosphere with cool spots. A few show very large amplitude excursions to the bright end within a single night, with little or no such behavior in V and I. We identify these as ``flare" variables and suspect that this may be an optical manifestation of the ubiquitous X-ray flares known to occur on T Tauri stars \citep{f05}. Finally, we have a set of stars which do not behave in either fashion and whose variability is likely linked to unsteady accretion or the rotation of a star with a hot (accretion) spot.

\begin{figure}
\plotone{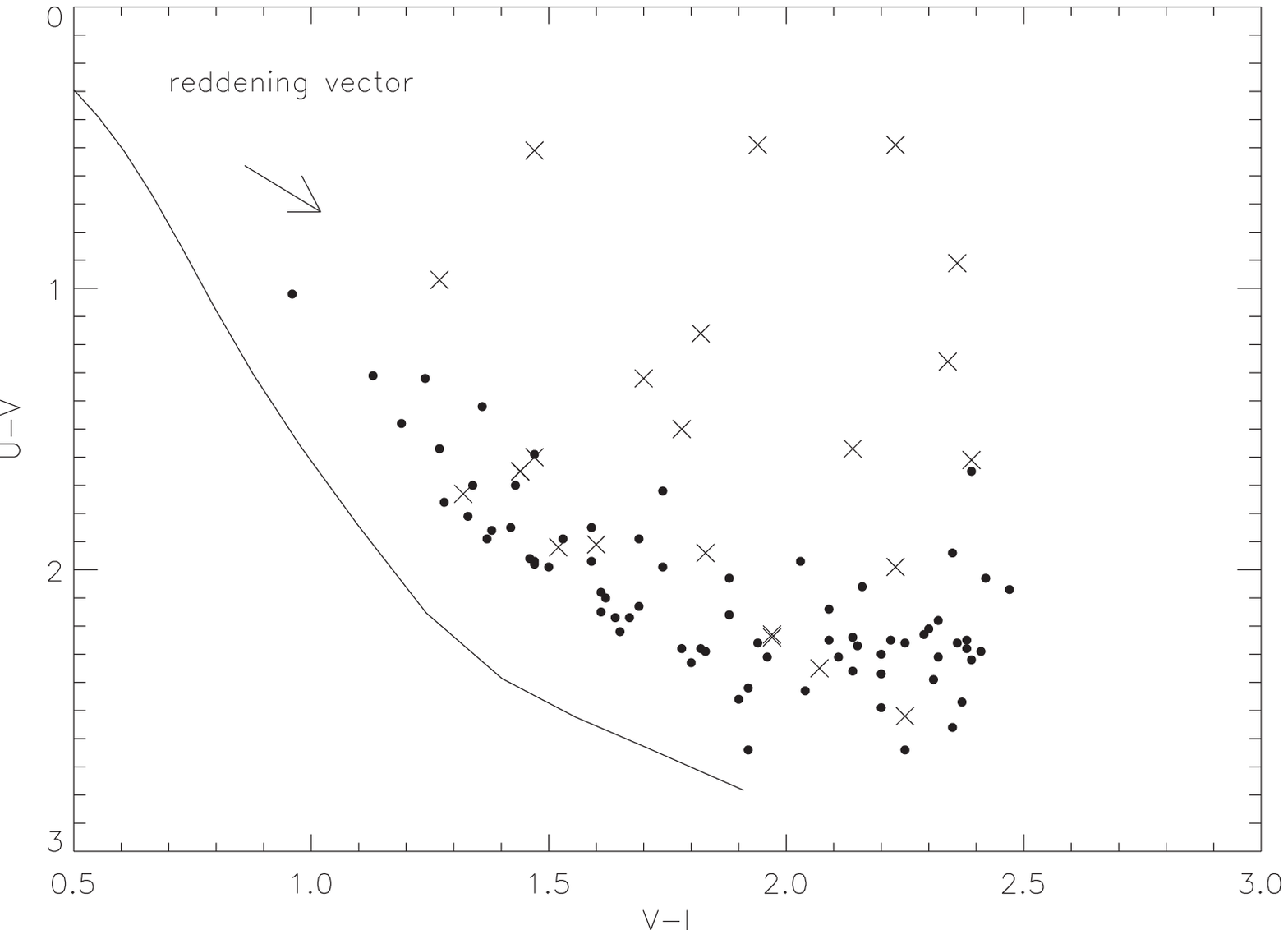}
\figcaption{Color-color diagram for periodic stars from \citet{m04} or \citet{l04} with R-I$<$1.3.  X's denote clearly variable stars in U based on our photometry. A color-color sequence from Bessell et al.\ (1998) for main sequence stars is shown along with a standard reddening vector. Stars scattering above the principal sequence have substantial U-V excesses.
\label{fig1}}
\end{figure}

 A color-color plot for the 95 stars in our field with known rotation periods and lying within the selected color range is shown in Fig. \ref{fig1}. Stars identified in this study as U-band variables are shown by crosses. Obviously, all of the stars on this plot must be U-band variables at some level since they have spotted surfaces and rotate. However, our study is focused on the larger amplitude variations that accompany accretion events and is not sensitive enough to detect some of the smaller variability associated with cool spots. It is clear from this figure that most of the stars in the periodic sample lie along a color-color line that approximates a reddened version of the standard main sequence color-color line, while some stars scatter above it. We define a U-V excess for all of the stars in our study by first determining their vertical displacement in this diagram from the reddened sequence of luminosity class V stars \citep{b98} and then subtracting 0.6 mag to normalize to the median of the non-variable sequence. The average U magnitude is used for variable stars. We adopted a value of E({\it B}-{\it V})=0.1 \citep{s97,p00,r02} as the foreground color excess for NGC 2264. This procedure leads to a rather clear separation between stars with significant U-V excess ($>$ 0.5 mag) and those without it.   

\subsection{U-V excess vs H$\alpha$ Equivalent Width}

It may reasonably be expected that the stars in Fig. \ref{fig1} with substantial U-V excesses are those which are still actively accreting from their circumstellar disks. This is supported by the fact that all but one of the U-V excess stars are also detected by us as variables. Since accretion is well known to be an unsteady process, we expect active accretors to be (irregular) variables, often of large amplitude. As a test of our presumption that both U-V excess and H$\alpha$ equivalent width are indicators of accretion activity, we show in Fig. \ref{fig2} the relationship between them. U-V excess are from this study and
H$\alpha$ equivalent widths are from \citet{ds05}. Note that we have plotted all of the stars in common to both studies here, not just the periodic ones. Clearly, the correlation is strong and that is confirmed by a Spearman Rank Order test which returns a probability of about 10$^{-30}$ that such a close association would occur by random. Since H$\alpha$ equivalent width is widely regarded as an accretion indicator and is even commonly used to distinguish WTTS from CTTS, we are gratified by this result.  

An interesting feature of this plot is that, while the overall correlation is excellent, when one looks at stars with equivalent widths of less than 10 \AA, the canonical division point for CTTS and WTTS, one still finds many stars with detectable U-V excesses. In fact, it is clear from Fig. \ref{fig2} that the correlation between U-V excess and H$\alpha$ equivalent width exists even for stars with equivalent widths less than 10 \AA, {\it albeit} with increased scatter. It is clear to us, based on this figure that not all stars classified as WTTS on the basis of their H$\alpha$ emission equivalent widths lack accretion disks. The same point has recently been made by \citet{khs06} based on their radiative transfer models of H$\alpha$ formation in CTTSs.

\subsection{U-V excess and Variability vs Rotation Period}

Armed with the evidence that our measured U-V excess indeed correlates with accretion, we now proceed to inquire whether it is also associated with rotation period. A plot of these quantities for our color-selected sample of stars with known rotation periods is shown in Fig. \ref{fig3}. Clearly, we do see an association between these two measured quantities, although it is not a one-to-one correlation. Also, clearly, the direction of the association is in the sense predicted by the disk-locking paradigm. Slowly rotating stars are more likely to have large U-V excesses than are rapidly rotating stars. 

To assign a confidence level to this result, we apply Fisher's Exact Test because our data closely approximate the conditions for its validity and because it is widely regarded as accurate and conservative even for relatively small sample sizes \citep{p01}. The test is appropriate to contingency tables of any size but we apply it here in the 2 $\times$ 2 form. Our data naturally divide themselves into the groups rapid rotators and slow rotators and the classes accretors and non-accretors. The only issue is choosing the dividing lines, which we do based on the marginal distributions, at a period of 6.3 days (log P = 0.8) and a U-V excess of 0.5 mag. Results are not sensitive to the exact choice of how to divide the data, as is evident from the plot.

The Fisher's Exact Test, like many statistical tests of its nature, comes in two forms: one-sided and two-sided. The one-sided form is appropriate when one is testing a prediction, as is the case here. The two-sided form is appropriate when one is seeking any correlation (i.e. in either direction) in the data. The prediction tested by the observational experiment reported here was that there would be a positive correlation between rotation period and U-V excess. The Fisher's Exact Test returns a p-value of 0.001 for this one-sided case, indicating that the hypothesis of association is confirmed with a confidence level of 99.9\%. 

It is interesting that the general form of the observed relationship is the same as that originally found in the ONC by \citet{e93} with near-infrared excess as the accretion disk proxy and the same as has been found in other such studies, as referenced above. Namely, we find that U-V excess is rare or absent among rapid rotators while it is more commonly present in slow rotators, although not ubiquitous. Rapid rotators with active accretion would be a challenge to the disk-locking paradigm so it is particularly significant that they are not seen. Slow rotators without active accretion, on the other hand, are not unexpected given the disk-locking paradigm. In this sample, they have at least two plausible interpretations. One possibility is that not all stars with accretion disks may be detectable as such using a particular proxy at a given time. Accretion is known to be unsteady on time scales of hours or days and may be unsteady on time scales of decades or centuries; we have no evidence on this point. Some of the slow rotators without U-V excess emission might well show such such excess in the future.

Another possibility is that the observed slow rotators without U-V excess may have simply lowered their accretion rates in the relatively recent past. The time scale on which accretion disappears is not well known but is likely to be short compared to the time scale on which a star can substantially change its rotation rate. In other words, one should not expect that the moment a star ceases to be substantially influenced by its disk, that it would immediately spin up. Indeed, the spin-up can only occur on the gravitational contraction time scale, which is of order a million years for stars of this mass and age range, while the disk dissipation time scale is probably an order of magnitude less. 

Another point to note about Fig. \ref{fig3} is that the dividing line between where active accretion is still found and where it is rare or absent is at about a period of 6 or 7 days (log P $\sim$ 0.8). This is consistent with the picture of rotational evolution presented recently by \citet{hm05}. They argue that the peaks in the rotation period distribution in NGC 2264 found by \citet{l04,l05} at 1 and 4 days are the spun-up counterparts of the well-known 2 and 8 day peaks seen in the ONC. This would suggest that stars with 4 day periods in NGC 2264 should not currently be locked to their disks and, indeed, that is what the data reported here also suggest. One may see on Fig. \ref{fig3} that the 4 day period stars (log P = 0.6) do not have many U-V excess examples, whereas among the 8 day period stars they are common.    

Finally, we mention the perhaps obvious point that some degree of natural variance should be expected among the stars in a sample like this. Eight days is a {\it representative} disk-locking period but it is certainly not the case that every star would be regulated to the same angular velocity. All disk-locking theories have parameters such as accretion rate, magnetic field strength and magnetic geometry that can affect the locking period. What the spread of locked angular velocities might be and what the time history of evolution of the effect might be are currently beyond our ability to know, either theoretically or empirically. The result reported here is that there is a statistically significant association between rotation and U-V excess in this sample that agrees with expectation based on the disk-locking paradigm.    

\section{Summary}

The observational experiment reported here was designed specifically to test a prediction of the disk-locking hypothesis, namely that slow rotation should be associated with active accretion disks in PMS stars. We used U-V excess as our accretion indicator following many previous studies in this area. We verified that there is an excellent correlation between our U-V excess measurements and H$\alpha$ equivalent widths reported by \citet{ds05}. We found, in fact, that the hypothesis tested is confirmed at about the 99.9\% level. This is significant new evidence in favor of the disk-locking paradigm, although it does not distinguish between various disk-locking theories or explain all the features of the rotational history of solar-like PMS stars that we would like to understand.

\begin{figure}
\plotone{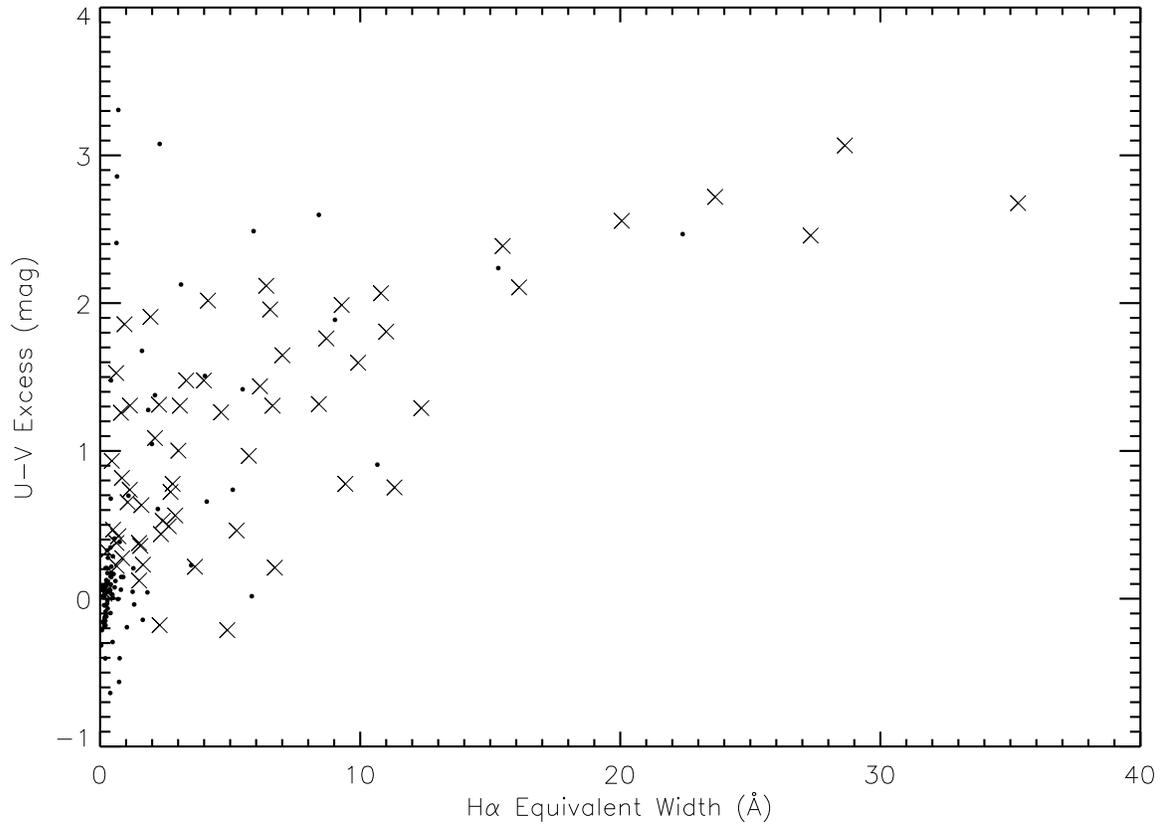}
\figcaption{U-V excess versus H$\alpha$ equivalent width from \citet{ds05} for stars in common to both data sets.  Variable stars are again represented by X's. The excellent correlation between the plotted quantities is apparent and is verified by a Spearman Rank Order test. \label{fig2}}
\end{figure}

\begin{figure}
\plotone{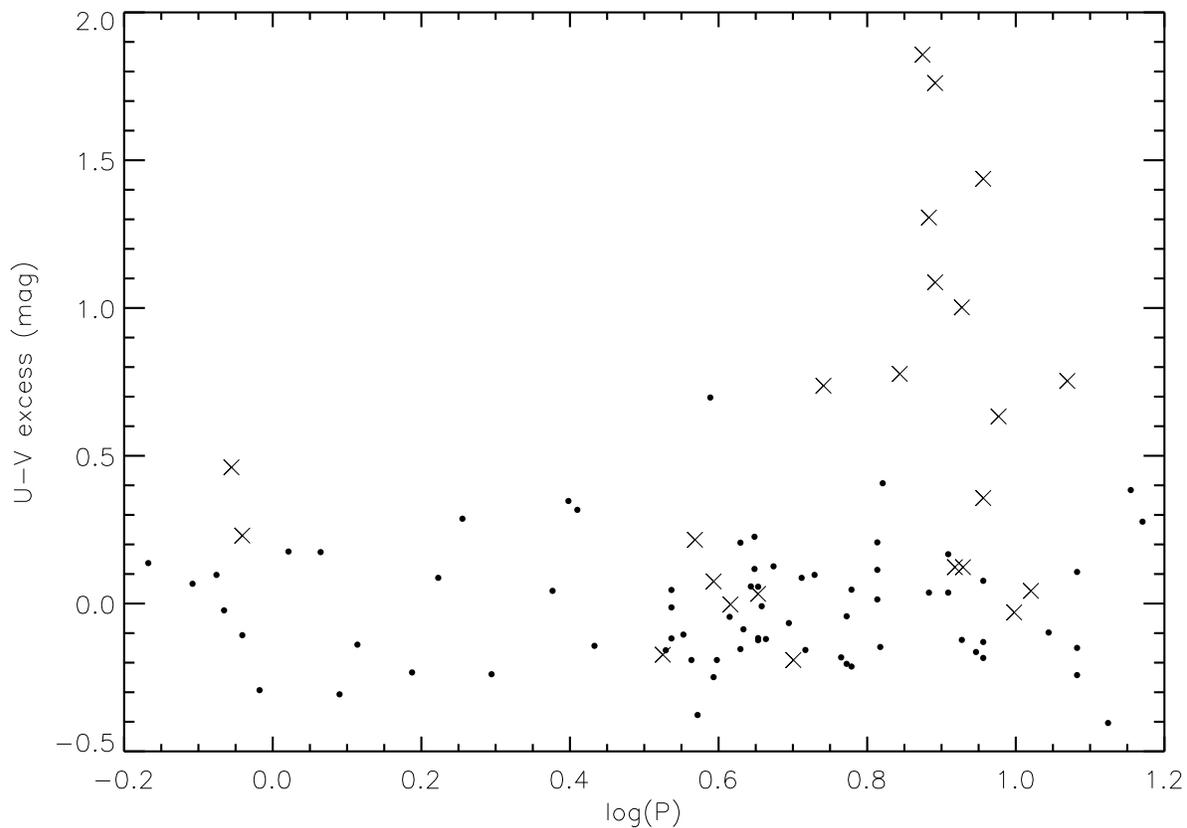}
\figcaption{U-V excess versus rotation period from \citet{m04} or \citet{l04} for higher mass members ({\it R}-{\it I} $<$ 1.3).   Again, X's represent variable stars, while filled circles are non-variable stars. A clear association between these variables is apparent and verified by a Fisher's Exact Test which returns p-values of around 0.001 for the one-sided case testing positive association when the division between rotation groups is placed at 6.3 days (log P = 0.8) and the division between accreting and non-accreting cases at 0.5 mag. \label{fig3}}
\end{figure}

Parenthetically, our study also supports the wide-spread belief that U-V excess (and, with some caution and caveats, variability) is an excellent indicator of accretion rates. Fig. \ref{fig2} suggests that U-V excess may even be more sensitive and useful than H$\alpha$ equivalent width at low accretion rates. Many stars that would be classified as WTTS by the H$\alpha$ criterion and thereby often assumed to be without accretion disks are actually still accreting, {\it albeit} at lower rates than their CTTS counterparts. U-band variability is also associated with accretion in the sense that every star but one with a U-V excess in our periodic sample is also a U-band variable. However, caution must be applied in using U-band variability alone to determine accretion status, since all PMS stars are probably variable in U at some level due to the effects of rotation, cool spots and flares unrelated to accretion events. 

Finally, we note that it will be interesting to compare how various accretion indicators correlate among themselves and with rotation period in other clusters and with expanded data sets. The two indicators for accretion used here -- U-V excess and H$\alpha$ equivalent width -- are both thought to arise on or close to the stellar surface from the accretion flow itself. The more commonly used proxy is infrared excess emission, but that is actually related to the presence of a reservoir for accretion, not to active accretion {\it per se}. Whether there is any difference in the degree of correlation between infrared excess and U-V excess (or H$\alpha$ equivalent width or velocity width) and rotation remains to be investigated. We plan to extend our U monitoring program to other extremely young clusters.

\acknowledgements

We gratefully acknowledge the contributions of J. Herbst to the analysis. This material is partly based on work supported by the National Aeronautics and Space Administration under Grant NAG5-12502 issued through the Origins of Solar Systems Program to W.H. We also thank Sigma Xi for its support of this work through a grant to C. F.

\clearpage

\end{document}